\documentclass[lettersize,journal]{IEEEtran}

\usepackage{amsmath,amsfonts}
\usepackage{algorithmic}
\usepackage[linesnumbered,ruled,vlined]{algorithm2e}
\usepackage{array}
\usepackage[caption=false,font=normalsize,labelfont=sf,textfont=sf]{subfig}
\usepackage{textcomp}
\usepackage{stfloats}
\usepackage{url}
\usepackage{verbatim}
\usepackage{graphicx}
\usepackage{multirow}
\usepackage{bm}
\usepackage{booktabs}
\usepackage[table]{xcolor}
\usepackage{cite}
\usepackage{longtable}
\usepackage{float}
\usepackage{amssymb}
\usepackage{listings}

\usepackage{tabularx}
\usepackage{tikz}
\usepackage{pgfplots}
\usepackage{calc}
\pgfplotsset{compat=1.18}

\newif\ifuseexternalbars
\useexternalbarsfalse

\newsavebox{\IssueTableBox}

\newlength{\CatTitleHeight}
\definecolor{RQSummaryBlue}{RGB}{226,239,250}
\newcommand{\RQSummary}[2]{%
    \par\medskip
    \noindent
    \begin{tikzpicture}
        \node[
            draw=black!25,
            fill=RQSummaryBlue,
            rounded corners=5pt,
            line width=0.4pt,
            inner xsep=10pt,
            inner ysep=8pt,
            text width=\dimexpr\linewidth-22pt\relax,
            align=justify
        ] {\textbf{Summary of #1:} #2};
    \end{tikzpicture}%
    \par\medskip
}

\usepackage[most]{tcolorbox}
\usetikzlibrary{shadows}
\tcbset{
  RQConBox/.style={
    enhanced,
    frame hidden,
    boxrule=0pt,
    arc=3pt, outer arc=3pt,
    colback=white, colbacklower=blue!3!white,
    left=8pt, right=6pt, top=4pt, bottom=4pt,
    borderline west={3pt}{0pt}{blue!60!black},
    drop shadow={shadow xshift=0.5pt, shadow yshift=-0.5pt, opacity=0.15},
    fontupper=\normalsize\rmfamily,
  }
}

\def\BibTeX{{\rm B\kern-.05em{\sc i\kern-.025em b}\kern-.08em
    T\kern-.1667em\lower.7ex\hbox{E}\kern-.125emX}}
\usepackage[hidelinks]{hyperref}

\begin{document}

\title{Enhancing the Non-Functional Quality Compliance of LLM-Generated Code through Quality-Aware Preference Learning}

\author{Liang~Lu, Yuan~Jiang\textsuperscript{*}, Christoph~Treude, Jingyu~Xiao, Xiaohong~Su, and~Michael~R.~Lyu%
\IEEEcompsocitemizethanks{
\IEEEcompsocthanksitem L. Lu (2021110360@stu.hit.edu.cn), Y. Jiang (jiangyuan@hit.edu.cn) and X. Su (sxh@hit.edu.cn) are with the School of Computer Science and Technology, Harbin Institute of Technology, Harbin, Heilongjiang, China. Liang Lu and Yuan Jiang contributed equally to this work, and Yuan Jiang is the corresponding author.\protect
\IEEEcompsocthanksitem C. Treude is with the School of Computing and Information Systems, Singapore Management University, Singapore. E-mail: ctreude@smu.edu.sg.\protect
\IEEEcompsocthanksitem J. Xiao (whalexiao99@gmail.com) and M.~R.~Lyu (lyu@cse.cuhk.edu.hk) are with the Department of Computer Science and Engineering, The Chinese University of Hong Kong, Hong Kong, China.\protect\\
}
}
\markboth{IEEE Transactions on Software Engineering}%
{Lu \MakeLowercase{\textit{et al.}}: Quality-Aware Preference Learning for LLM-Generated Code}

\maketitle

\begin{abstract}
Large Language Models (LLMs) have been widely adopted in commercial code completion engines, significantly enhancing coding efficiency and productivity. However, even functionally correct LLM-generated code may exhibit non-functional quality issues that violate coding standards and best practices, such as poor style and limited maintainability.
To address this, we propose a framework for quality-aware preference learning that guides LLMs toward generating criteria-compliant code. Our approach consists of three phases. First, we construct a dataset of paired criteria-violating and criteria-compliant samples, where each pair contains code exhibiting a specific non-functional quality issue and its repaired version that resolves the issue. Second, we design an adaptive token weighting mechanism to emphasize quality-sensitive code regions.
Third, we introduce a hybrid optimization objective that combines ranking loss with language modeling loss and KL divergence to enable effective comparative optimization.
Extensive experiments on DeepSeek-Coder and Qwen2.5-Coder show that our method substantially improves compliance with the targeted non-functional quality criteria while maintaining functional correctness, achieving a 75.7\% relative increase in Quality Reciprocal Score (QRS) on MBPP-sanitized for Qwen2.5-Coder. Fine-tuning a 7B model requires less than three hours, indicating strong practical viability. Ablation studies and a user study further support the effectiveness of the proposed framework.
\end{abstract}


\section{Introduction}
\IEEEPARstart{L}{arge} Language Models (LLMs) have revolutionized various fields with their impressive capabilities in understanding and generation~\cite{brown2020language, chowdhery2022palm}. In software engineering, LLMs now support a wide range of code generation scenarios, from code completion to agent-based and autonomous development, significantly improving productivity~\cite{chen2021evaluating, roziere2023code,peng2023impact,liang2024large,zhang2024autocoderover}. For example, interactive assistant tools (e.g., GitHub Copilot~\cite{copilot2021github} and Amazon CodeWhisperer~\cite{amazon2023codewhisperer}) support real-time code authoring, while agent-based frameworks (e.g., AutoGPT~\cite{autogpt2023_autogpt}, SGAgent~\cite{zhang2026sgagent} and SWE-Agent~\cite{yang2024swe}) enable multi-step program synthesis and automated issue resolution.
Recent reports from Microsoft indicate that GitHub Copilot has reached over 20 million users, with extensive adoption across enterprise environments~\cite{microsoft2025annualreport}.

Despite these benefits, LLMs often generate code with non-functional quality issues. In this study, we focus specifically on code quality issues that can be identified by widely used static analysis tools, including code smells and violations of established coding conventions and best practices that affect readability, maintainability, and code structure~\cite{velasco2025causal,liu2024refining}. Recent studies report that even functionally correct code generated by models such as ChatGPT frequently exhibits such issues, with 53\% of Java code and 37\% of Python code showing style and maintainability problems~\cite{liu2024refining}. This suggests that functionally correct code may still require additional developer effort to address these detectable quality issues, potentially reducing the productivity gains offered by LLM-based code generation.

Enhancing the non-functional quality of LLM-generated code is therefore essential for reducing technical debt and improving developer productivity, and is critical to the long-term success of AI-driven code generation. Although prior work has investigated non-functional quality issues using static analysis tools~\cite{siddiq2022empirical,siddiq2024quality,liu2024refining}, relatively few studies have focused on improving models' intrinsic ability 
to directly generate code that better adheres to established non-functional quality criteria.

Existing training paradigms for improving the non-functional code quality of LLM-generated code exhibit several limitations. Specifically, (1) \textit{Instruct tuning}~\cite{li2024instructcoder,muennighoff2023octopack}, which fine-tunes models on high-quality code, lacks contrastive signals between good and poor examples, making it difficult for models to internalize quality preferences. (2) \textit{Prompt-based mitigation} methods~\cite{velasco2025causal,della2025prompt,tony2025prompting}, which inject handcrafted instructions (e.g., ``avoid code smell'') into prompts, often fail due to the model’s limited ability to distinguish quality among semantically equivalent forms. (3) \textit{Prefix-based specialization} methods such as SVEN~\cite{he2023large, he2024instruction}, which learn separate positive and negative prefixes to guide generation, rely on single-sample conditioning and lack explicit contrastive supervision, thus limiting their generalizability. (4) \textit{Alignment methods} like DPO~\cite{rafailov2023direct}, while effective in natural language tasks, treat all tokens equally and ignore the structural salience of quality-sensitive regions, making them suboptimal for quality-aware code generation. These representative methods are adopted as baselines to benchmark the effectiveness of our proposed approach.

To address these limitations, we propose a novel framework for quality-aware preference learning. First, we construct a paired dataset in which each sample consists of a criteria-compliant variant and a corresponding criteria-violating counterpart, enabling direct supervision of non-functional quality differences across diverse issues.
Second, we introduce adaptive token weights that dynamically prioritize quality-sensitive regions during training. Tokens associated with high uncertainty receive greater emphasis, while well-learned or neutral tokens are down-weighted, allowing the model to focus its capacity more effectively.
Third, we leverage these paired samples to define a sequence-level ranking loss that encourages the model to prefer criteria-compliant code over its criteria-violating alternative. This ranking objective is integrated with language modeling and Kullback–Leibler divergence objectives, creating a hybrid optimization objective that enables effective comparative optimization while maintaining the model's generation capability. Through this framework, we examine whether non-functional quality preferences derived from widely adopted coding criteria can be effectively learned by code LLMs and generalized across diverse programming tasks.

We conduct experiments on DeepSeek-Coder~\cite{guo2024deepseek} and Qwen2.5-Coder~\cite{hui2024qwen2}, demonstrating substantial improvements in compliance with the targeted non-functional quality criteria, including a 75.7\% increase in the Quality Reciprocal Score (QRS) on MBPP-sanitized for Qwen2.5-Coder.  
Our approach also outperforms the evaluated baseline methods, achieving  
a 45.0\% improvement in QRS compared to DPO on CodeContests for Qwen2.5-Coder while maintaining functional correctness. In addition, a user study shows that developers consistently prefer outputs from models optimized with our method. Ablation studies further analyze the contribution of the individual components in our framework.

The main contributions are as follows.

\begin{itemize}
    \item We present a novel comparative optimization framework for improving LLM-generated code compliance with non-functional quality criteria.
    
    \item We design a data construction pipeline tailored for learning non-functional quality preferences, addressing the lack of suitable resources.
    
    \item We propose a ranking loss with adaptive token weights that emphasizes quality-sensitive tokens during training. 
    
    \item We conduct comprehensive experiments across two code LLMs and three benchmarks, demonstrating consistent improvements in targeted quality-criteria compliance while preserving functional correctness.
    
\end{itemize}

\section{Motivation Example}
\label{background}
\begin{figure}[htbp] 
    \centering 
    \includegraphics[width=1.0\linewidth]{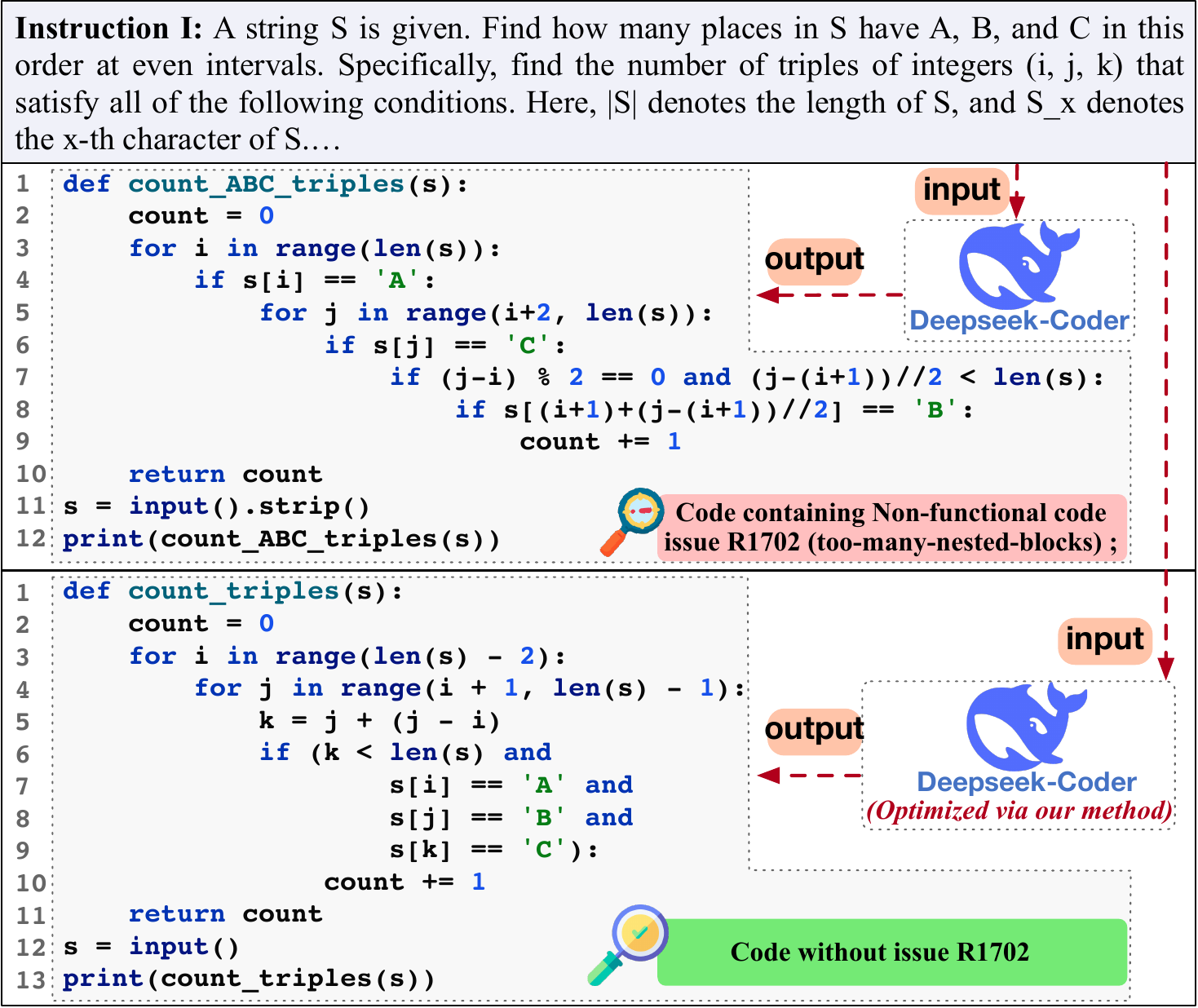} \caption{Comparison of original and optimized DeepSeek-Coder outputs on a LiveCodeBench task. The original contains issue R1702 (too-many-nested-blocks), while the optimized output via our method avoids this issue.} 
    
    \label{fig:code_example} 
\end{figure}

Code LLMs may produce outputs that satisfy functional requirements but frequently exhibit non-functional quality issues~\cite{liu2024refining}. To illustrate this, we examine a representative task from the \textit{LiveCodeBench} dataset~\cite{jain2024livecodebench} (top of Fig.\ref{fig:code_example}). When the task instruction is fed to the publicly released DeepSeek-Coder model~\cite{guo2024deepseek}, the generated solution passes all test cases but contains a non-functional issue, \texttt{R1702: too-many-nested-blocks}, caused by excessive nesting of loops and conditionals (see top half of Fig.~\ref{fig:code_example}).

The presence of deeply nested control structures in LLM-generated code obscures core semantic logic and increases cognitive complexity. More broadly, such patterns reflect a common challenge in code generation, as they degrade readability and maintainability by requiring developers to track multiple control-flow contexts~\cite{campbell2018cognitive}. These issues are especially problematic in collaborative or production settings, where clarity and long-term maintainability are critical.

This example highlights the importance of addressing non-functional quality issues in LLM-generated code. Although these issues do not affect functional correctness, they significantly impact developer productivity and long-term software quality. To address this, we propose a quality-aware preference learning framework to guide LLMs toward criteria-compliant code that adheres to established structural and stylistic conventions. As shown in the bottom half of Fig.~\ref{fig:code_example}, our optimized model produces functionally correct code without the R1702 violation, illustrating the potential of our approach.

\section{Method Overview}
\label{framework}

\begin{figure*}[htbp]
    \centering
    \includegraphics[width=\linewidth]{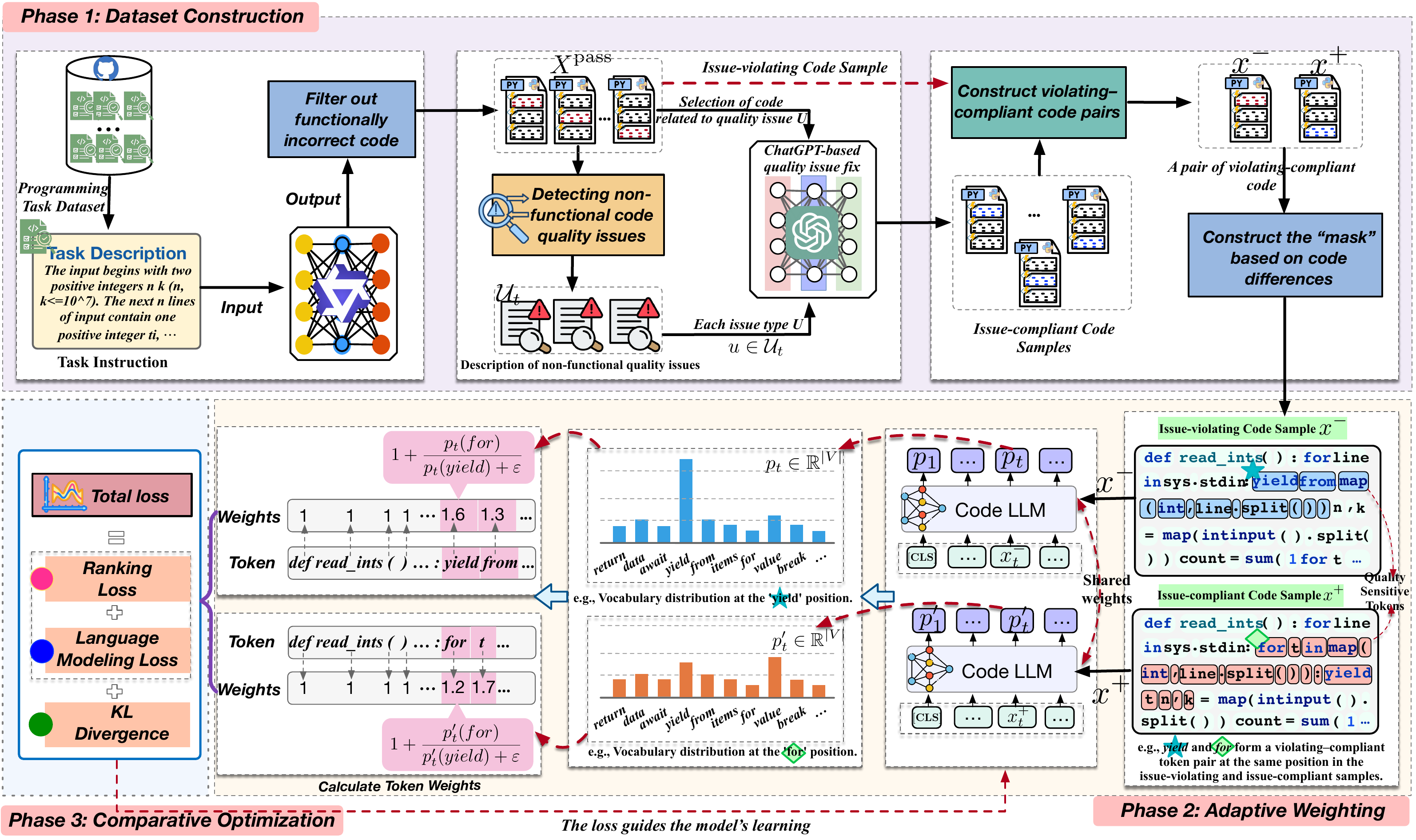}
    \caption{Workflow of our non-functional quality-aware code generation framework based on violating–compliant supervision. The pipeline includes dataset construction, token-level weighting using LLM-predicted distributions, and multi-loss optimization to guide the model toward criteria-compliant generation.}
    \label{fig:framework}
\end{figure*}

Our framework, as illustrated in Fig.~\ref{fig:framework}, enhances quality-aware code generation in three key stages.

\subsection{Dataset Construction for Criteria-violating and criteria-compliant Code}
We begin by constructing a dataset of paired criteria-violating and criteria-compliant code samples. Given a task description, we use a target LLM (e.g., DeepSeek-Coder) to generate candidate solutions. To ensure functional correctness, only code samples that pass all test cases are retained. These samples are then analyzed for non-functional quality issues. For each quality type, we apply automated repair to the original code exhibiting that issue, generating a corresponding criteria-compliant version. By aligning the original and repaired code, we extract token-level differences at corresponding positions, forming a set of violating–compliant token pairs.

\subsection{Adaptive Weighting for Quality-Sensitive Tokens}
For each violating–compliant token pair, we compute token-specific weights based on the model's predicted probability distributions. Specifically, we obtain the vocabulary-level distribution at the differing token position when feeding the criteria-violating code into the target LLM ($ \mathbf{p}_t $), and similarly for the criteria-compliant code ($\mathbf{p}'_t$). These distributions are used to derive a weighting scheme that encourages the model to prefer tokens from the criteria-compliant variant while penalizing those from the criteria-violating variant. This weighting mechanism provides fine-grained supervision at the token level.

\subsection{Comparative Optimization for Quality-Aware Preference Learning}
To incorporate the preference supervision induced by paired criteria-compliant and criteria-violating code samples into model training, we introduce three complementary loss functions. First, a sequence-level ranking loss \( \mathcal{L}_{\text{rank}} \) promotes preference for criteria-compliant code over criteria-violating code for the same task. Second, a KL-divergence loss \( \mathcal{L}_{\text{KL}} \) ensures that the model's output remains close to the original LLM in regions unrelated to code quality. Third, a language modeling loss \( \mathcal{L}_{\text{LM}} \) preserves functional correctness by maintaining alignment with the original generation behavior. Together, these losses guide the model toward generating high-quality, compliant code while retaining task-level correctness.

\section{Details of the Proposed Method}
\label{our_method}

\subsection{Dataset Construction}
\label{sec:data_construction}
High-quality datasets are crucial for effectively training models to produce code with improved non-functional quality. However, existing datasets often lack precisely labeled examples and token-level annotations for non-functional code issues. To overcome these limitations, we construct a new dataset by collecting paired samples of criteria-violating and compliant code for each specific type of quality issue.

\subsubsection{Detection of Non-Functional Quality Issues}
\label{sec:static_detector}
Most existing studies assess the non-functional quality of LLM-generated code using static analysis tools~\cite{siddiq2022empirical}. For example, ~\cite{liu2024refining,siddiq2022empirical,siddiq2024quality} and \cite{rabbi2024ai} employ \textsc{Pylint}~\cite{pylint}, Feng et al.~\cite{feng2023investigating} use \textsc{Flake8}~\cite{ziade27flake8}, and Yetiştiren et al.~\cite{yeticstiren2023evaluating} choose \textsc{SonarQube}~\cite{sonarqube2024}. To ensure broader coverage and complementarity, we integrate all three tools in our evaluation. 
Following the common categorization practices used in real-world software development (as reflected in \textsc{Pylint}), we classify the quality issues detected by these tools into four categories: \textbf{E} (Error; severe programming issues such as potential bugs that may affect program execution), \textbf{R} (Refactor; structural problems that diverge from established best practices), \textbf{C} (Convention; violations of coding standards, including naming and formatting), and \textbf{W} (Warning; stylistic or minor programming issues that impact readability and maintainability). We systematically review all available detection rules from the three tools and identify a total of 368 rules from \textsc{Pylint} (E=126, W=133, C=50, R=59), 129 from \textsc{Flake8} (E=31, W=23, C=73, R=2), and 232 from \textsc{SonarQube} (E=48, W=61, C=72, R=51).
To avoid redundant detections, we apply each rule to representative code examples and compare violation positions and messages across tools to identify equivalent rules. We deduplicate overlapping rules and consolidate them into a unified set of 527 distinct rules: E=142, W=160, C=128, and R=97.
This rule set is fixed before model training and operationalizes the non-functional quality criteria targeted in this study. These criteria capture statically detectable aspects of code quality and are not intended to constitute an exhaustive definition of software quality.

\subsubsection{Dataset Construction Workflow}

Algorithm~1 presents our dataset construction process, consisting of three main steps: program generation, quality issue detection, and issue repair.
For program generation, we begin with 5{,}000 training tasks from the APPS dataset~\cite{hendrycks2021measuring}. For each task, we use its natural language description as the input $I$, and prompt the target model (e.g., DeepSeek-Coder 7B) with $I$ to generate $n{=}100$ code samples, following the hyperparameter settings detailed in Section~\ref{subsec:implementation} (Lines~3–4). Next, we retain only those generated samples that pass all functional test cases, forming a set $X^{\text{pass}}$ of candidate programs that satisfy functional correctness while still exhibiting diverse non-functional quality issues (Line 5).
For quality issue detection, we perform static analysis on the functionally correct candidates $X^{\text{pass}}$ using the unified rule set $\mathcal{R}$ (Lines~6). For each type of quality issue, we select up to 50 code samples from $X^{\text{pass}}$ that are flagged by the detection tools as exhibiting that issue (Lines~8-9). 

For the issue repair process, we employ an advanced ChatGPT model (i.e., GPT\textendash 4) as a rule-guided transformation engine to convert each criteria-violating sample into a \emph{criteria-compliant} variant that addresses only the targeted quality issue.
To preserve the original coding structure, we enforce minimal, issue-local edits by constraining the repair model to modify only the statements flagged by the detector, and then re-check the transformed code with the same static analyzers and the test suite to ensure that the target issue is fully eliminated while functional correctness is preserved, retaining only the sample that is verified to be criteria-compliant and functionally correct (Lines~10–11).
This process yields a pair of criteria-violating and criteria-compliant code samples, denoted as $(x^{-}, x^{+})$. We identify edited lines between $x^{-}$ and $x^{+}$ using \texttt{difflib.ndiff} and use tokenizer offsets to mark the tokens in these lines, producing separate mask vectors $(m^{-}, m^{+})$. Each element in the mask is set to $1$ if the token participates in the rule-triggering or rule-fixing region, and $0$ otherwise. Finally, we assemble the training instance $(I, x^{+}, m^{+}, x^{-}, m^{-})$ and add it to the quality dataset $\mathcal{D}_{\text{quality}}$ (Lines~12--13).
Note that although ChatGPT can fix non-functional quality issues when given the exact issue type and location, this post-hoc repair process requires iterative detection and correction steps, and cannot ensure that all quality issues are resolved. It is also more costly, less predictable, and harder to scale. In contrast, we aim to internalize quality preferences during training, enabling the model to generate high-quality code by default and avoid repeated generate-detect-repair loops.

\begin{algorithm}[ht]
    \small
    \caption{Construction Pipeline for Dataset $\mathcal{D}_{\text{quality}}$}
    \label{alg:dataconstruction}
    \KwIn{APPS training tasks $\mathcal{T}$;\\
        \hspace*{3.2em}Quality detector $\mathcal{Q}$ (wrapper of \textsc{Pylint}, \textsc{Flake8}, \textsc{SonarQube} with unified rule set)}
    \KwOut{Dataset $\mathcal{D}_{\text{quality}}$}
    
    Initialize dataset $\mathcal{D}_{\text{quality}} \leftarrow \emptyset$\;
    
    \For{each task $t \in \mathcal{T}$}{
        Extract the natural-language task instruction $I$\;
        Generate $n$ code samples $\{x_i\}_{i=1}^n$ using the target LLM\;
        
        Filter out samples that fail \emph{any} test case and retain those passing all tests to form $X^{\text{pass}}$\;
        
        Perform quality analysis on $X^{\text{pass}}$ using detector $\mathcal{Q}$\;
        Let $\mathcal{U}_t$ be the set of all quality issues detected in $X^{\text{pass}}$\;   
        
        \For{each issue type $u \in \mathcal{U}_t$}{    
            $S_u \leftarrow$ select at most 50 samples in $X^{\text{pass}}$ that violate issue type $u$\;
            
            \For{each $x^{-} \in S_u$}{
                $x^{+} \leftarrow$ fix issue $u$ for $x^{-}$ via the ChatGPT under a minimal-edit constraint, and re-check with $\mathcal{Q}$ and the test suite to ensure both the elimination of $u$ and functional correctness\;
                
                Construct token-level mask vectors $(m^{-}, m^{+})$ identifying the quality-relevant token changes between $x^{-}$ and $x^{+}$\;
                
                Add instance $(I, x^{+}, m^{+}, x^{-}, m^{-})$ to dataset $\mathcal{D}_{\text{quality}}$\;
            }
        }
    }
\end{algorithm}

\subsection{Adaptive Weighting for Quality-Sensitive Tokens}

Building on the dataset constructed in the previous subsection, we next describe how the paired criteria-violating and criteria-compliant programs are used to provide fine-grained supervision during training. Existing preference optimization approaches, such as Direct Preference Optimization (DPO)~\cite{rafailov2023direct}, generally compare preferences at the sequence level without explicitly distinguishing the contributions of individual tokens. However, in our setting, only a small subset of tokens is directly associated with the presence or absence of a particular quality issue, while most tokens mainly provide shared syntactic and contextual information. Treating all tokens equally may therefore dilute the supervision associated with quality-related modifications. To address this limitation, we introduce an adaptive token-level weighting scheme that assigns greater importance to quality-sensitive positions.

Formally, let $x^{+}$ and $x^{-}$ denote the criteria-compliant and criteria-violating code sequences, respectively. During dataset construction, \texttt{difflib.ndiff} identifies the edited regions between the two programs, and tokenizer offsets are used to mark the corresponding quality-sensitive token positions. Within these regions, we perform position-wise comparison according to the autoregressive generation position. Specifically, $(y_i^{+}, y_i^{-})$ denotes the tokens produced at generation position $i$ in the compliant and violating sequences, respectively.

This position-wise comparison should not be interpreted as semantic token alignment between $x^{+}$ and $x^{-}$. Instead, it compares the generation choices made by the compliant and violating trajectories at the same autoregressive step. This distinction is particularly clear at the first divergent position $i^{*}$, where $x^{+}_{<i^{*}} = x^{-}_{<i^{*}}$ but $x^{+}_{i^{*}} \neq x^{-}_{i^{*}}$. At this position, the two tokens are competing alternatives under exactly the same preceding context. Increasing the preference for $x^{+}_{i^{*}}$ therefore directly encourages the model to follow the compliant generation trajectory. Since subsequent token distributions depend on previously generated tokens, an earlier preference can also influence later predictions~\cite{liang2025edit}.

After the two sequences diverge, their preceding contexts may differ, particularly for repairs involving insertions or deletions. We nevertheless preserve the original autoregressive positions rather than shifting tokens according to lexical alignment, because the probability of each token is determined by the prefix available at its actual generation step. Thus, the purpose of the position-wise comparison is to steer the generation trajectory rather than to establish semantic correspondence between later tokens. If one sequence has no token available at position $i$, the unavailable cross-token probability is set to zero.

For a quality-sensitive position $i$, we obtain the next-token probability distribution from the current model. The adaptive weight $w_i^{+}$ for the compliant token is defined as
\begin{equation}
    w_i^{+}
    =
    1 + \mathrm{clip}\!\left(
    \frac{
        P_{\theta}(y_i^{-} \mid x_{<i}^{+}, I)
    }{
        P_{\theta}(y_i^{+} \mid x_{<i}^{+}, I) + \varepsilon
    },
    0, 2
    \right),
    \label{eq:token-weight}
\end{equation}
where $\varepsilon>0$ ensures numerical stability, and the ratio is clipped to $[0,2]$ to avoid excessively large weights.

Intuitively, the numerator measures how strongly the model still favors the violating alternative $y_i^{-}$ under the compliant autoregressive context, while the denominator measures its confidence in the compliant token $y_i^{+}$. If the violating alternative remains competitive, the ratio increases and the corresponding compliant token receives a larger weight. Conversely, when the model already strongly prefers the compliant token, $w_i^{+}$ approaches $1$. The adaptive weight for a quality-sensitive token in $x^{-}$ is defined analogously using the corresponding compliant alternative under the violating context.

During dataset construction, the mask vectors $(m^{+},m^{-})$ are binary and indicate whether each token belongs to a quality-sensitive region. During training, we replace the binary value at each quality-sensitive position with its adaptive weight:
\begin{equation}
    \begin{aligned}
        m_i^{+}
        &=
        \begin{cases}
            w_i^{+}, & \text{if position } i \text{ is quality-sensitive},\\
            1, & \text{otherwise},
        \end{cases}\\[3pt]
        m_i^{-}
        &=
        \begin{cases}
            w_i^{-}, & \text{if position } i \text{ is quality-sensitive},\\
            1, & \text{otherwise}.
        \end{cases}
    \end{aligned}
    \label{eq:mask-vector}
\end{equation}

The quality-sensitive positions are fixed by the masks constructed from the repaired code pairs, whereas their adaptive weights are recomputed from the current model predictions during training. The computed weights are detached during backpropagation, so they reweight token-level log-likelihood terms without introducing an additional gradient path through the weighting procedure.

This mechanism provides two complementary benefits. First, it places additional supervision on tokens directly associated with quality-related modifications. Second, it adjusts the strength of this supervision according to the model's current preference between compliant and violating generation choices. Consequently, positions at which the model remains uncertain or continues to favor undesirable patterns receive stronger optimization signals.

\subsection{Comparative Optimization for Quality-Aware Preference Learning}

\begin{figure}[htbp]
    \centering
    \includegraphics[width=1.0\linewidth]{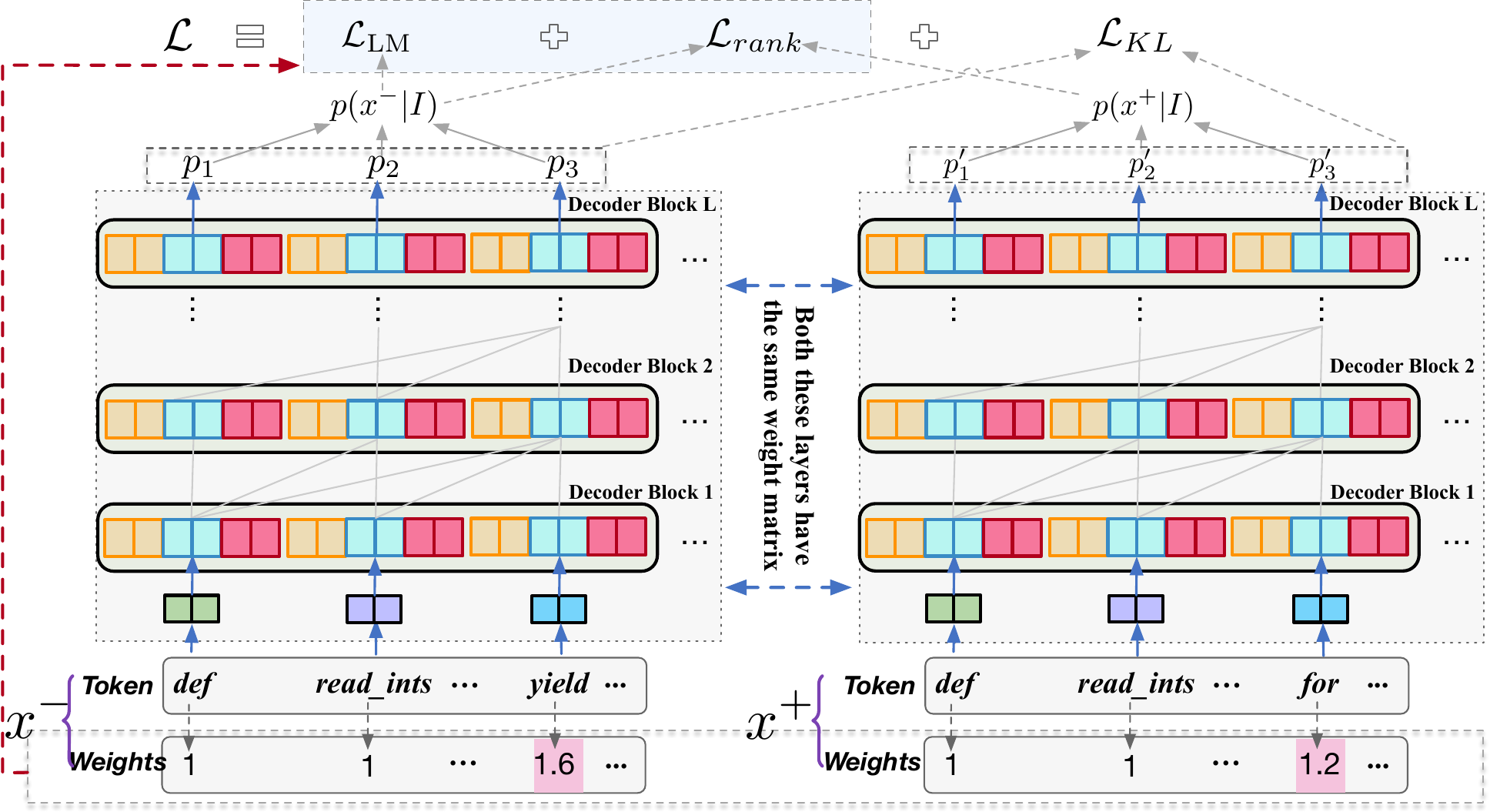}
    \caption{Overview of the comparative optimization framework for quality-aware preference learning, integrating the ranking loss, language modeling loss, and KL-divergence loss.}
    \label{fig:comparative_learning}
\end{figure}

We next introduce our comparative optimization objective, which integrates the ranking loss with adaptive token weights, the language modeling loss, and the KL-divergence loss, as illustrated in Fig.~\ref{fig:comparative_learning}. Together, these objectives improve non-functional code quality while constraining the optimized model to retain its original code-generation capability.

\subsubsection{Ranking Loss with Adaptive Token Weights}

The ranking loss encourages the model to assign a higher likelihood to the criteria-compliant code $x^{+}$ than to its criteria-violating counterpart $x^{-}$. Following prior sequence-level calibration and preference optimization methods based on cumulative autoregressive log-likelihoods~\cite{zhao2023calibrating,rafailov2023direct}, we use the cumulative sequence log-likelihood as the basic preference signal and incorporate the adaptive token weights defined above to emphasize quality-sensitive positions.

This formulation is well suited to our paired dataset because each $x^{+}$ is obtained through a localized repair of the corresponding functionally correct $x^{-}$. Specifically, 63.0\% of repairs modify fewer than 10\% of the tokens, and 82.8\% modify fewer than 20\% (Section~\ref{subsec:dataset}). Thus, most of the program context remains shared between each pair, while the adaptive weights focus the optimization on the regions responsible for their quality differences. We also evaluated a length-normalized variant in preliminary experiments, but it yielded lower overall performance than the cumulative formulation adopted here.

Formally, given a code sequence $x$, its adaptive weight vector $m$, the task description $I$, and model parameters $\theta$, we define the weighted sequence score as
\begin{equation}
    s_{\text{mask}}(x,m,I)
    =
    \sum_{t=1}^{|x|}
    m_t
    \log P_{\theta}
    \left(
        x_t \mid x_{<t}, I
    \right).
    \label{eq:masked-score}
\end{equation}

Each sequence is evaluated according to its own autoregressive context. The adaptive weights determine how strongly individual token log-likelihoods contribute to the sequence score, with additional emphasis placed on quality-sensitive positions where the model has not yet clearly distinguished compliant from violating generation choices.

We then define the sequence-level ranking loss as
\begin{equation}
    \mathcal{L}_{\mathrm{rank}}
    =
    -\log \sigma\left(
        s_{\text{mask}}(x^{+},m^{+},I)
        -
        s_{\text{mask}}(x^{-},m^{-},I)
    \right),
    \label{eq:masked-ranking-loss}
\end{equation}
where $\sigma(\cdot)$ denotes the sigmoid function.

Minimizing $\mathcal{L}_{\mathrm{rank}}$ increases the model's relative preference for the criteria-compliant sequence while assigning stronger learning signals to quality-sensitive positions. In this way, the objective captures both the sequence-level preference between $x^{+}$ and $x^{-}$ and the localized token-level differences responsible for their non-functional quality distinction.

\subsubsection{Language Modeling Loss}
To keep the model aligned with standard code generation practices while emphasizing high-quality examples, we apply the language modeling loss \emph{only} to the criteria-compliant code $x^{+}$. Moreover, we incorporate adaptive token weights during loss computation to focus the optimization on quality-sensitive tokens. Thus,
\begin{equation}
    \mathcal{L}_{\text{LM}} = 
    - \sum_{t=1}^{T} m_{t}^{+} \,\log P\bigl(x_{t}^{+} \mid x_{<t}^{+}, I\bigr)
\end{equation}
where $T$ is the length of $x^{+}$. By assigning higher weights to quality-sensitive positions, we reinforce the model's tendency to produce superior patterns where quality matters most.

\subsubsection{Loss Term for Preserving Functional Correctness}

To ensure that our optimized model preserves functional correctness when improving non-functional code quality, we introduce a $\mathcal{L}_{\text{KL}}$ loss term, which computes the KL divergence between the next-token distributions of the optimized model $P(x_t \mid x_{<t}, I)$ and the original LLM $P_{\text{ref}}(x_t \mid x_{<t}, I)$: 
\begin{equation}
    \mathcal{L}_{\text{KL}}
    =
    \sum_{t\in\mathcal{N}}
    \mathrm{KL}\!\left(
    P_{\theta}(\cdot \mid x_{<t}, I)
    \,\|\, 
    P_{\text{ref}}(\cdot \mid x_{<t}, I)
    \right),
\end{equation}
where $\mathcal{N}$ denotes the set of quality-insensitive token positions, i.e., tokens that are not involved in quality-related modifications. By applying $\mathcal{L}_{\text{KL}}$ only to these unchanged regions, we constrain the optimized model to remain close to the original LLM in its functional behavior.

\subsubsection{Overall Training Loss}

The overall objective, which combines the above loss terms for quality-aware preference learning, is defined as follows:

\begin{equation}
    \mathcal{L} = w_1 \mathcal{L}_{\text{LM}} + w_2 \mathcal{L}_{\text{rank}} + w_3 \mathcal{L}_{\text{KL}},
\end{equation}

where $w_1$, $w_2$, and $w_3$ are weighting coefficients that balance the contributions of $\mathcal{L}_{\text{LM}}$, $\mathcal{L}_{\text{rank}}$, and $\mathcal{L}_{\text{KL}}$, respectively.

\subsection{Advantages of Our Method}
To the best of our knowledge, this work is among the first to explicitly target the improvement of non-functional quality in LLM-generated code, aiming to promote better adherence to coding standards and best practices. Our approach offers the following key advantages: (1) it introduces a comparative optimization framework with a ranking loss that explicitly guides the model to prefer criteria-compliant code over criteria-violating alternatives, thereby improving non-functional code quality; (2) it incorporates adaptive token weighting to dynamically emphasize uncertain and quality-sensitive regions, enabling more effective quality-aware preference learning; and (3) it adopts a multi-loss training objective that combines \(\mathcal{L}_{\mathrm{LM}}\), \(\mathcal{L}_{\mathrm{rank}}\) and \(\mathcal{L}_{\mathrm{KL}}\), ensuring that quality improvements are achieved without compromising functional correctness.

\section{Experimental Design}
\label{sec:experiments}

\subsection{Research Questions}

We aim to answer the following research questions (RQs)\footnote{The source code and evaluation scripts for this study are publicly available at~\url{https://github.com/luliang9966/LLM_CodeQuality_Enhance_TSE_2026}.}:

RQ1: To what extent does our method improve the non-functional quality of generated code compared to the original models, while maintaining functional correctness?

RQ2: To what extent does our method outperform representative existing methods in enhancing non-functional code quality?

RQ3: How do the individual components of our method (e.g., ranking loss) and variations affect performance?

RQ4: To what extent do users prefer code from our optimized model over the baseline?

\subsection{Evaluation Metrics}
\label{subsec:evaluation_metrics}
We evaluate our method using metrics for non-functional code quality and functional correctness.


\textbf{Pass Rate}: 
This metric measures non-functional code quality by the proportion of samples that contain no detected non-functional quality issues among all generated samples. The pass rate has been widely adopted in prior studies as an effective indicator of overall code quality~\cite{yao2025training}. 
Formally, given a set of generated programs $\mathcal{P}$, the pass rate is defined as:
\begin{equation}
    \label{eq:pass_rate_definition}
    \begin{aligned}
        \mathrm{Pass \ Rate}(\mathcal{P})
        &= \frac{1}{|\mathcal{P}|}
        \sum_{p \in \mathcal{P}}
        \mathbb{I}\!\left[
        \substack{p\ \text{contains no detected}\\[-1pt]
        \text{quality issues}}
        \right],
    \end{aligned}
\end{equation}
where $\mathbb{I}[\cdot]$ is the indicator function. This metric reflects the practical usability of generated code, indicating the fraction of outputs that  contain no detected violations of the targeted static-analysis criteria.



\textbf{Quality Reciprocal Score (QRS):}
This metric quantifies the quality-issue burden of a complete generated program by considering the number and severity of detected violations. Let $\mathrm{QI}(p)$ denote the weighted number of quality issues detected in program $p$. We define QRS as:
\begin{equation}
    \label{eq:qrs_definition}
    \mathrm{QRS}(p) = \frac{1}{\mathrm{QI}(p) + 1}.
\end{equation}
A program with fewer or less severe quality issues receives a higher QRS, while $\mathrm{QRS}(p)=1$ indicates that no quality issues are detected. The reciprocal formulation provides a continuous program-level measure that distinguishes programs with different quality-issue burdens while reducing the influence of programs containing a large number of violations on aggregate results. Following the severity treatment adopted by \textsc{Pylint}~\cite{pylint}, each \textbf{E}-type issue is assigned a weight of five, while other issue types retain unit weight.

\textbf{Rationale.}
Pass Rate and QRS capture complementary aspects of non-functional code quality. Pass Rate measures how frequently a model generates programs without any detected non-functional quality issues, providing an intuitive indicator of issue-free generation. In contrast, QRS provides a continuous measure of the quality-issue burden of a complete program, allowing improvements to be captured even when some violations remain.

Since an LLM may generate multiple solutions per prompt, we adopt a task-level evaluation strategy. For each task, we compute the Pass Rate and the mean QRS over all generated solutions, and then average these statistics across tasks within a given dataset.


\textbf{Pass@k}:
Although the primary objective of our method is to improve non-functional code quality, maintaining functional correctness remains essential. Therefore,
we use the Pass@k metric~\cite{chen2021evaluating} to evaluate functional correctness. For each task, we generate $k$ code samples and check whether any of them pass all the unit tests provided in the evaluated dataset. The Pass@k metric is defined as the percentage of tasks for which at least one of the $k$ generated samples is functionally correct.

\subsection{Dataset}
\label{subsec:dataset}

\begin{figure}[htbp]
    \centering
    \includegraphics[width=\linewidth]{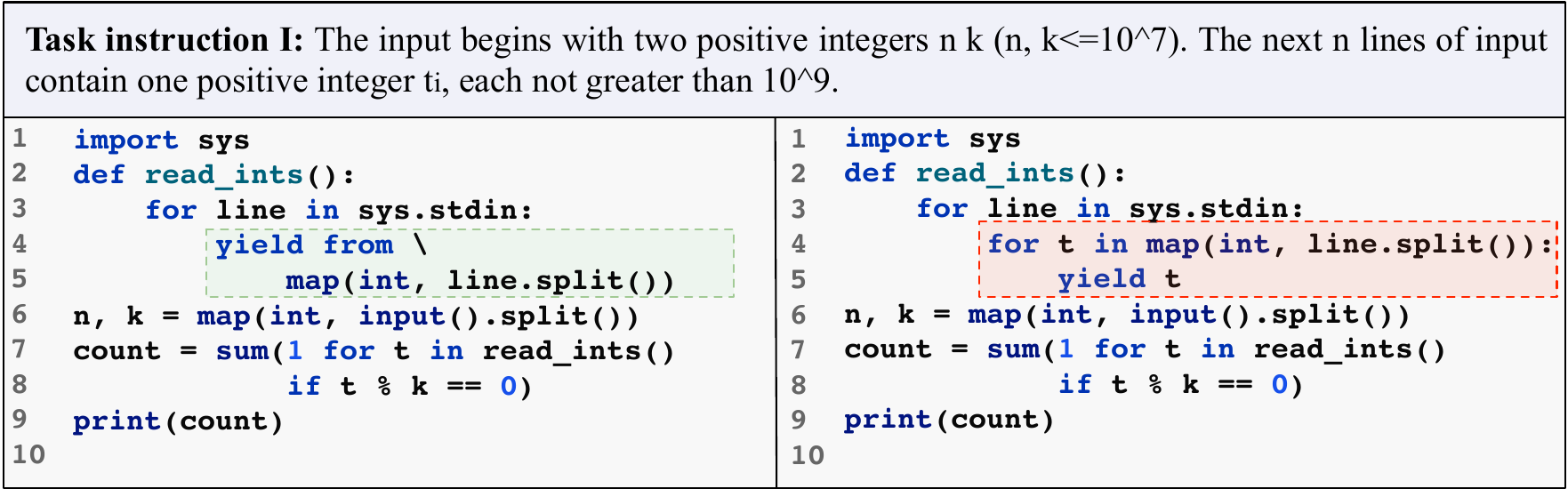}
    \caption{A data instance including instruction $I$, criteria-compliant code $x^+$ (bottom left), criteria-violating code $x^-$ (bottom right), and their corresponding mask vectors $m^+$ and $m^-$. Tokens with mask value 1 (highlighted) correspond to quality-sensitive regions; others are assigned 0.}
    \label{fig:example_sample}
\end{figure}

After applying Algorithm~\ref{alg:dataconstruction}, we obtain 3{,}106 valid criteria-compliant and criteria-violating code pairs covering 164 non-functional quality issues. 
To substantiate the high structural similarity within each pair, we quantify the locality of repair edits by measuring the token-level edit ratio between the violating and compliant versions. The statistics indicate that the modifications are highly localized: 63.0\% of the repairs affect fewer than 10\% of tokens, and 82.8\% affect fewer than 20\%, indicating that most criteria-compliant variants are derived from small, targeted changes rather than extensive rewrites. Only 5.0\% of the pairs exceed a 50\% edit ratio, which is expected, as a small fraction of quality issues are inherently structural and necessitate more substantial refactoring.
Therefore, the constructed pairs cover a diverse range of programming tasks and maintain both clear quality contrasts and high structural similarity, thus providing strong learning signals for training the target model to generate higher-quality code.
An example data instance from our dataset is shown in Fig.~\ref{fig:example_sample}. Each instance consists of the task instruction $I$, the criteria-compliant variant $x^{+}$, the criteria-violating sample $x^{-}$, and their corresponding mask vectors $m^{+}$ and $m^{-}$.

To assess whether our method can effectively enhance the non-functional code quality of LLM-generated programs, we evaluate the fine-tuned models on several widely used code generation benchmarks. Following prior work~\cite{gao2025seer}, we adopt \textit{MBPP-sanitized}~\cite{austin2021program}, \textit{LiveCodeBench}~\cite{jain2024livecodebench}, and \textit{CodeContests}~\cite{li2022competition} as our evaluation datasets. We deliberately exclude simpler benchmarks such as \textit{HumanEval}~\cite{chen2021evaluating}, since the generated solutions for these tasks are typically short and thus less likely to contain non-functional quality issues.

\subsection{Evaluation Setting}
\label{subsec:implementation}
For our experiments, we select DeepSeek-Coder~\cite{guo2024deepseek} (specifically, the 7B version DeepSeek-Coder-Instruct-v1.5) and Qwen2.5-Coder~\cite{hui2024qwen2} (specifically, the 7B version Qwen2.5-Coder-Instruct), two open-source code LLMs that have demonstrated strong performance on code generation tasks and are widely used for evaluating code generation enhancement methods~\cite{gao2025seer}. In addition, both models offer a favorable balance between size and performance, making them well-suited for experimentation under typical computational constraints.

We implement our method using PyTorch and the Hugging Face Transformers library~\cite{wolf2020transformers}, and employ LoRA for
parameter-efficient fine-tuning with rank $r=32$ and scaling factor
$\alpha=16$. Training is conducted on NVIDIA A6000 GPUs, each equipped with 48 GB of memory. The total training time is approximately 3 hours, showcasing the efficiency of our method in terms of computational resources.
During training, we employ a learning rate of 1e-4, a batch size of 1, and the AdamW optimizer~\cite{loshchilov2018decoupled}. 
To tune the loss coefficients, we evaluated $w_1,w_2,w_3 \in \{1,2,3,4,5\}$ on a held-out validation split and selected the configuration that best balanced QRS and functional correctness, without using the evaluation benchmarks. This resulted in $(w_1,w_2,w_3)=(1,1,2)$ for DeepSeek-Coder and $(1,4,2)$ for Qwen2.5-Coder.
To stabilize the training process, gradient clipping is applied with a maximum norm of 1.0.
During evaluation, following~\cite{he2023large}, we use random sampling with a temperature of 0.4 and a top-p of 0.95 to generate 30 solutions per task.

\section{Experiment Results}
\label{eval_results}
\subsection{RQ1: To what extent does our method improve the non-functional quality of generated code compared to the original models, while maintaining functional correctness?}
\label{sec:rq1}

To answer this question, we evaluate the original DeepSeek-Coder~\cite{guo2024deepseek} and Qwen2.5-Coder~\cite{hui2024qwen2} models and their corresponding optimized versions produced by our method. We conduct experiments on three benchmarks, namely \textit{MBPP-sanitized}, \textit{LiveCodeBench}, and \textit{CodeContests}. Non-functional code quality is evaluated using QRS and Pass Rate, while functional correctness is measured using Pass@k~\cite{chen2021evaluating}.

\paragraph{Non-Functional Code Quality}
Table~\ref{tab:rq1_code_quality} reports the QRS and Pass Rate of the original and optimized models across the three benchmarks. The optimized models consistently achieve higher scores for both metrics. For example, on \textit{MBPP-sanitized}, Qwen2.5-Coder achieves a QRS of 79.4\% after optimization, compared with 45.2\% for the original model, corresponding to a relative improvement of 75.7\%. Its Pass Rate also increases from 18.5\% to 64.6\%, indicating a substantial increase in the proportion of generated programs without detected non-functional quality issues. 
Similar improvements are observed for DeepSeek-Coder and across the other benchmarks, showing a consistent improvement pattern across models and datasets.

\begin{table}[ht]
    \centering
    \caption{Non-functional code quality of the original and optimized models across MBPP-sanitized, LiveCodeBench, and CodeContests.}
    \label{tab:rq1_code_quality}
    \resizebox{\linewidth}{!}{%
        \begin{tabular}{lcccccc}
            \toprule
            \multirow{2}{*}{\textbf{Approach}}
            & \multicolumn{2}{c}{\textbf{MBPP-sanitized}}
            & \multicolumn{2}{c}{\textbf{LiveCodeBench}}
            & \multicolumn{2}{c}{\textbf{CodeContests}} \\
            \cmidrule(lr){2-3}\cmidrule(lr){4-5}\cmidrule(lr){6-7}
            & \textbf{QRS} & \textbf{Pass Rate}
            & \textbf{QRS} & \textbf{Pass Rate}
            & \textbf{QRS} & \textbf{Pass Rate} \\
            \midrule
            \multicolumn{7}{c}{DeepSeek-Coder} \\
            \midrule
            Original  & 45.3\% & 3.3\%  & 40.4\% & 19.8\% & 37.9\% & 11.7\% \\
            Optimized & \textbf{67.9\%} & \textbf{39.6\%} & \textbf{71.8\%} & \textbf{56.8\%} & \textbf{79.5\%} & \textbf{67.5\%} \\
            \midrule
            \multicolumn{7}{c}{Qwen2.5-Coder} \\
            \midrule
            Original  & 45.2\% & 18.5\% & 42.0\% & 26.0\% & 34.2\% & 8.6\% \\
            Optimized & \textbf{79.4\%} & \textbf{64.6\%} & \textbf{66.3\%} & \textbf{46.1\%} & \textbf{68.6\%} & \textbf{48.8\%} \\
            \bottomrule
        \end{tabular}%
    }
\end{table}

To assess the statistical significance of these improvements, we perform paired bootstrap resampling at the task level. Across the six model--benchmark comparisons, the 95\% BCa confidence intervals for QRS improvements range from $[+20.88, +24.29]$ to $[+39.73, +43.59]$, while those for Pass Rate improvements range from $[+18.41, +21.85]$ to $[+52.04, +59.43]$. All confidence intervals remain above zero, and all improvements are statistically significant after Holm correction (adjusted bootstrap $p<0.001$), providing statistical support for the consistent quality gains observed across models and benchmarks.


To further characterize the improvement beyond aggregate QRS and Pass Rate, we examine the number of detected quality issues on a per-task basis. For each task, we generate 30 candidate solutions using DeepSeek-Coder~\cite{guo2024deepseek} and select the five solutions with the highest QRS scores. We then aggregate the quality issues detected in these five solutions using the static analysis tools described in Section~\ref{sec:static_detector}. A lower issue count indicates better non-functional quality.

\begin{figure}[htbp]
    \centering
    \includegraphics[width=\linewidth]{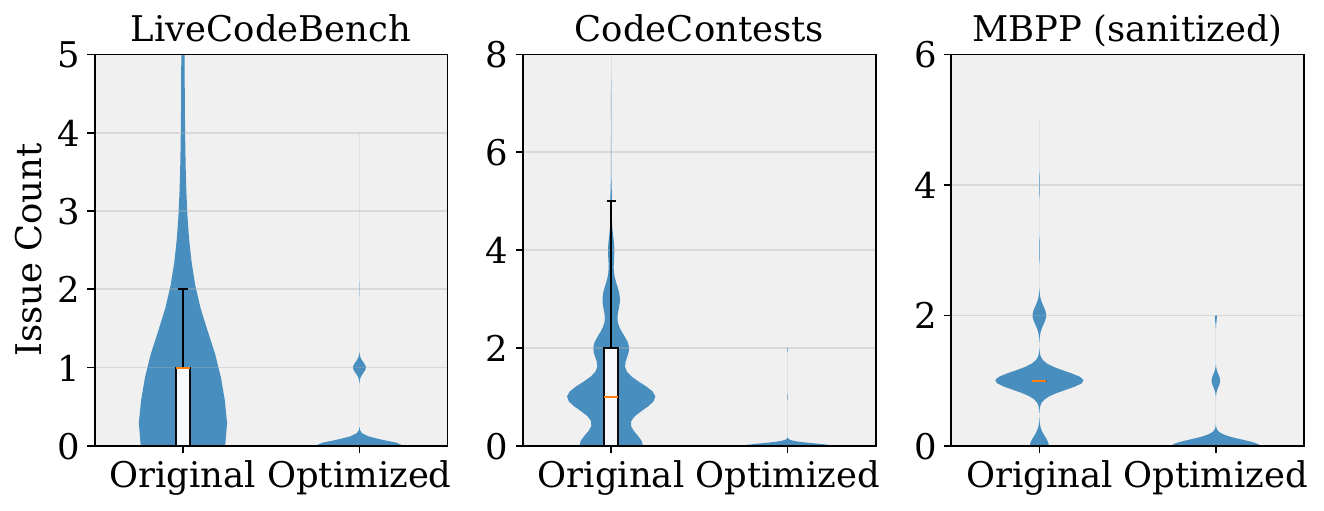}
    \caption{Distribution of detected quality issues per task for the original and optimized DeepSeek-Coder models. Each point represents the aggregated number of issues in the five solutions with the highest QRS selected from 30 candidates for a task.}
    \label{fig:violin_issue_distribution}
\end{figure}

Fig.~\ref{fig:violin_issue_distribution} shows the resulting issue-count distributions. Across all three benchmarks, the optimized model produces fewer detected quality issues than the original model. This result is consistent with the improvements in QRS and Pass Rate and provides additional evidence that our method reduces non-functional quality issues across programming tasks.

\begin{figure*}[htbp]
    \centering
    \includegraphics[width=\linewidth]{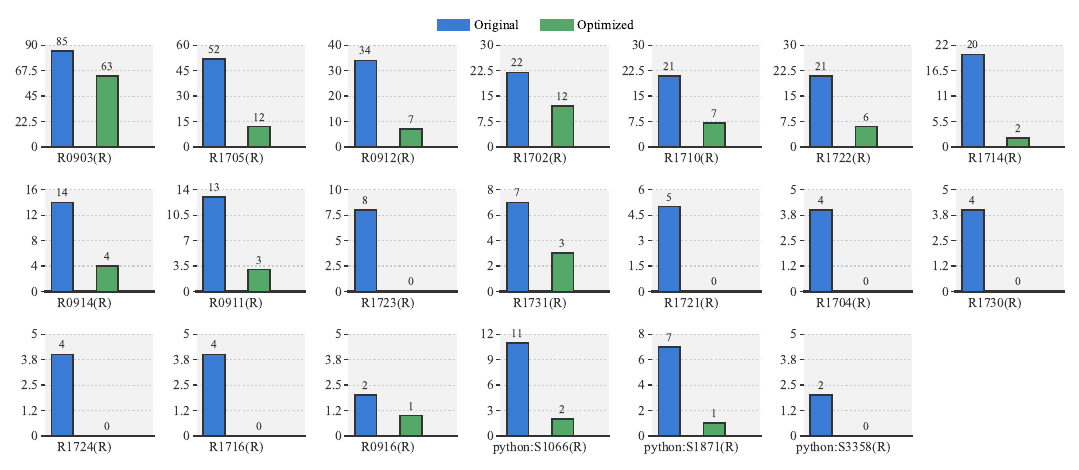}
    \caption{Occurrences of individual \textit{Refactor} (R-category) issues before and after optimization on \textit{LiveCodeBench}. Each bar pair reports the total occurrences across the five solutions with the highest QRS selected from the 30 candidates generated for each task by the original and optimized DeepSeek-Coder models. The x-axis denotes the issue IDs reported by the detection tools.}
    \label{fig:issuewise_reduction}
\end{figure*}

We further analyze \textit{Refactor} (R-category) issues, which capture structural and maintainability concerns such as excessive branching, deeply nested control flow, and inconsistent return structures. For each \textit{LiveCodeBench}~\cite{jain2024livecodebench} task, we select the five solutions with the highest QRS from the 30 candidates generated by the original and optimized DeepSeek-Coder models and aggregate the occurrences of individual R-category issues. As shown in Fig.~\ref{fig:issuewise_reduction}, the optimized model reduces a broad range of these issues. For example, R0912 (\textit{too-many-branches}) decreases from 34 to 7 occurrences, corresponding to a reduction of 79.4\%, while R1705 (\textit{no-else-return}) decreases from 52 to 12 occurrences, corresponding to a reduction of 76.9\%. These results indicate that the improvements are not limited to aggregate quality scores but extend to specific structural and maintainability issues.

\paragraph{Functional Correctness}
We next examine whether the improvement in non-functional quality affects functional correctness. Table~\ref{tab:rq1_functional_correctness} reports Pass@5 and Pass@10 across the three benchmarks. DeepSeek-Coder maintains or improves functional correctness after optimization on all three datasets. For Qwen2.5-Coder, the results also remain comparable to or improve over those of the original model, except for a small decrease on \textit{LiveCodeBench}. For example, its Pass@5 decreases slightly from 17.9\% to 17.4\%, while Pass@10 decreases from 20.3\% to 19.9\%. Overall, these results indicate that the substantial gains in non-functional code quality are achieved without a systematic degradation in functional correctness.

\begin{table}[ht]
    \centering
    \caption{Functional correctness of the original and optimized models across the three benchmarks.}
    \label{tab:rq1_functional_correctness}
    \resizebox{\linewidth}{!}{%
        \begin{tabular}{lcccccc}
            \toprule
            \multirow{2}{*}{\textbf{Approach}}
            & \multicolumn{2}{c}{\textbf{MBPP-sanitized}}
            & \multicolumn{2}{c}{\textbf{LiveCodeBench}}
            & \multicolumn{2}{c}{\textbf{CodeContests}} \\
            \cmidrule(lr){2-3}\cmidrule(lr){4-5}\cmidrule(lr){6-7}
            & \textbf{Pass@5} & \textbf{Pass@10}
            & \textbf{Pass@5} & \textbf{Pass@10}
            & \textbf{Pass@5} & \textbf{Pass@10} \\
            \midrule
            \multicolumn{7}{c}{DeepSeek-Coder} \\
            \midrule
            Original  & 64.4\% & 67.6\% & 14.3\% & \textbf{16.7\%} & 14.5\% & 18.9\% \\
            Optimized & \textbf{65.6\%} & \textbf{68.7\%} & \textbf{14.4\%} & \textbf{16.7\%} & \textbf{17.7\%} & \textbf{21.6\%} \\
            \midrule
            \multicolumn{7}{c}{Qwen2.5-Coder} \\
            \midrule
            Original  & 70.8\% & 73.7\% & \textbf{17.9\%} & \textbf{20.3\%} & 17.6\% & 22.2\% \\
            Optimized & \textbf{72.1\%} & \textbf{74.0\%} & 17.4\% & 19.9\% & \textbf{18.7\%} & \textbf{22.4\%} \\
            \bottomrule
        \end{tabular}%
    }
\end{table}

\RQSummary{RQ1}{Our method consistently improves the non-functional quality of LLM-generated code across models and benchmarks, with QRS gains of up to 75.7\% and substantially more issue-free outputs, while maintaining functional correctness.}

\subsection{RQ2: To what extent does our method outperform representative existing methods in enhancing non-functional code quality?}
\label{rq2}

To answer this question, we compare our method with four baseline approaches for improving the non-functional quality of code LLMs using our constructed datasets. The baselines are:
(1) Instruct Tuning, which fine-tunes on criteria-compliant samples to learn high-quality coding patterns, but relies solely on supervised learning. (2) PromptMit~\cite{velasco2025causal}, a prompt-based mitigation approach that steers the model away from non-functional quality issues by appending the phrase ``without introducing any code smell'' to the input prompt. Importantly, we avoid explicitly listing all issue types in prompts, as this would require long, brittle, and tool-specific instructions to be repeatedly injected at inference time, thereby increasing prompt length, cost, and maintenance burden in real-world use.
(3) SVEN~\cite{he2023large}, a preference-based method originally proposed for enhancing code security, which we adapt for non-functional quality improvement by training it on our dataset.
(4) DPO~\cite{rafailov2023direct}, a preference optimization approach that learns from paired examples of higher- and lower-quality code while treating all tokens equally. To ensure a fair comparison, all learning-based methods are trained on the same constructed dataset, and all methods (including PromptMit) are evaluated using the same unified set of 527 non-functional quality rules (see Section~\ref{sec:static_detector}) derived from three static analyzers.

\paragraph{Code Quality Comparison}
Table~\ref{tab:rq2_comparison} reports the QRS and Pass Rate of our method and baselines across the three benchmarks. For example, when optimizing Qwen2.5-Coder on \textit{CodeContests}, DPO~\cite{rafailov2023direct}, a representative preference optimization baseline, achieves 47.3\% QRS and 22.4\% Pass Rate, whereas our method achieves 68.6\% and 48.8\%, respectively. This corresponds to a relative QRS improvement of approximately 45.0\% over DPO. Similar improvements are observed across the other models and benchmarks.
Unlike DPO, which applies preference optimization at the sequence level without explicitly distinguishing quality-sensitive tokens, our method incorporates adaptive token weighting into the ranking objective to emphasize tokens associated with non-functional quality differences. Together with the comparative supervision provided by criteria-compliant and criteria-violating pairs, this design enables the optimization to focus more directly on quality-relevant distinctions. The contribution of each component is further examined in the ablation study in RQ3. To assess the statistical significance of these improvements, we perform paired bootstrap resampling at the task level between our method and DPO. Across the six model--benchmark comparisons, the 95\% BCa confidence intervals for QRS improvements range from $[+0.63, +3.35]$ to $[+19.23, +23.09]$, while those for Pass Rate improvements range from $[+0.67, +5.59]$ to $[+22.95, +29.72]$. All confidence intervals exclude zero, and the differences remain statistically significant after Holm correction (adjusted bootstrap $p\leq0.013$), providing statistical support for the consistent improvements over DPO.


\begin{table}[ht]
    \centering
    \caption{Comparison of non-functional code quality across models optimized by different methods.}
    \label{tab:rq2_comparison}
    \resizebox{\linewidth}{!}{%
        \begin{tabular}{lcccccc}
            \toprule
            \multirow{2}{*}{\textbf{Approach}}
            & \multicolumn{2}{c}{\textbf{MBPP-sanitized}}
            & \multicolumn{2}{c}{\textbf{LiveCodeBench}}
            & \multicolumn{2}{c}{\textbf{CodeContests}} \\
            \cmidrule(lr){2-3}\cmidrule(lr){4-5}\cmidrule(lr){6-7}
            & \textbf{QRS} & \textbf{Pass Rate}
            & \textbf{QRS} & \textbf{Pass Rate}
            & \textbf{QRS} & \textbf{Pass Rate} \\
            \midrule
            \multicolumn{7}{c}{DeepSeek-Coder} \\
            \midrule
            Instruct Tuning & 48.2\% & 5.8\% & 40.5\% & 10.5\% & 40.8\% & 5.4\% \\
            PromptMit  & 51.9\% & 17.0\% & 45.9\% & 40.3\% & 37.5\% & 16.3\% \\
            SVEN       & 66.4\% & 37.2\% & 61.5\% & 42.1\% & 58.4\% & 39.3\% \\
            DPO        & 65.9\% & 36.5\% & 68.0\% & 49.7\% & 66.5\% & 48.1\% \\
            Ours       & \textbf{67.9\%} & \textbf{39.6\%} & \textbf{71.8\%} & \textbf{56.8\%} & \textbf{79.5\%} & \textbf{67.5\%} \\
            \midrule
            \multicolumn{7}{c}{Qwen2.5-Coder} \\
            \midrule
            Instruct Tuning & 45.9\% & 2.1\% & 39.9\% & 7.8\% & 39.5\% & 3.3\% \\
            PromptMit  & 48.9\% & 7.3\%  & 37.4\% & 22.1\% & 31.6\% & 7.4\% \\
            SVEN       & 63.3\% & 33.3\% & 51.1\% & 27.9\% & 46.3\% & 23.9\% \\
            DPO        & 62.9\% & 49.0\% & 48.9\% & 33.1\% & 47.3\% & 22.4\% \\
            Ours       & \textbf{79.4\%} & \textbf{64.6\%} & \textbf{66.3\%} & \textbf{46.1\%} & \textbf{68.6\%} & \textbf{48.8\%} \\
            \bottomrule
        \end{tabular}%
    }
\end{table}

\paragraph{Functional Correctness}
Table~\ref{tab:rq2_functional_correctness} presents the Pass@k results across the three datasets, comparing our method against the baselines. Overall, our method maintains functional correctness at levels comparable to or slightly exceeding those of the baselines. For example, on \textit{CodeContests}, our method achieves a Pass@5 score of 17.7\%, compared to 12.6\% for DPO, corresponding to a relative improvement of 40.5\%. This demonstrates that carefully balancing \(\mathcal{L}_{\mathrm{LM}}\), \(\mathcal{L}_{\mathrm{rank}}\), and \(\mathcal{L}_{\mathrm{KL}}\) not only improves non-functional code quality but also preserves functional correctness. Instruct Tuning occasionally achieves slightly higher correctness scores by directly optimizing the language modeling loss, while providing limited improvement in non-functional quality. A similar pattern is observed for PromptMit, which retains the original model parameters and thus avoids potential accuracy degradation, but also results in lower non-functional quality.

\begin{table}[ht]
    \centering
    \caption{Functional correctness comparison of our method and four baselines, evaluated via Pass@k.}
    \label{tab:rq2_functional_correctness}
    \resizebox{\linewidth}{!}{%
        \begin{tabular}{lcccccc}
            \toprule
            \multirow{2}{*}{\textbf{Approach}}
            & \multicolumn{2}{c}{\textbf{MBPP-sanitized}}
            & \multicolumn{2}{c}{\textbf{LiveCodeBench}}
            & \multicolumn{2}{c}{\textbf{CodeContests}} \\
            \cmidrule(lr){2-3}\cmidrule(lr){4-5}\cmidrule(lr){6-7}
            & \textbf{Pass@5} & \textbf{Pass@10} & \textbf{Pass@5} & \textbf{Pass@10} & \textbf{Pass@5} & \textbf{Pass@10} \\
            \midrule
            \multicolumn{7}{c}{DeepSeek-Coder} \\
            \midrule
            Instruct Tuning & 62.7\%  &  65.1\% &  13.7\%  & 15.1\%  &  17.1\%  & 20.2\%   \\
            PromptMit  & 62.3\% & 64.9\% & \textbf{14.8\%} & 16.6\% & 12.7\% & 16.9\% \\
            SVEN & 64.6\% & 68.4\% & 12.6\% & 15.2\% & 10.0\% & 14.1\% \\
            DPO  & 65.0\% & 68.4\% & 12.6\% & 15.3\% & 12.6\% & 17.1\% \\
            Ours & \textbf{65.6\%} & \textbf{68.7\%} & 14.4\% & \textbf{16.7\%} & \textbf{17.7\%} & \textbf{21.6\%} \\
            \midrule
            \multicolumn{7}{c}{Qwen2.5-Coder} \\
            \midrule
            Instruct Tuning &  71.2\% &  72.7\%  &  \textbf{17.5\%}  & 19.1\%  & \textbf{19.3\%}   & \textbf{22.4\%} \\
            PromptMit  & 71.3\% & 73.4\% & 16.3\% & 18.5\% & 15.8\% & 20.0\% \\
            SVEN & 71.2\% & \textbf{74.0\%} & 14.6\% & 17.1\% &  13.9\% & 18.2\%\\
            DPO  & 68.2\% & 72.9\% & 15.9\% & 18.7\% & 16.5\% & 20.9\% \\
            Ours & \textbf{72.1\%} & \textbf{74.0\%} & 17.4\% & \textbf{19.9\%} & 18.7\% & \textbf{22.4\%} \\
            \bottomrule
        \end{tabular}%
    }
\end{table}
\paragraph{Efficiency Considerations} Table~\ref{tab:efficiency_comparison} compares the training efficiency of different methods when applied to DeepSeek-Coder on our dataset. As shown in the table, our method requires similar resources and training time as DPO and SVEN (approx. 49 minutes/epoch on a single A6000 GPU), while enabling significant non-functional quality gains. PromptMit incurs no training cost but suffers from inferior performance. Instruct tuning, which uses only criteria-compliant data (i.e., half of our dataset), yields the shortest training time but leads to the weakest quality performance.
During inference, all methods require the full model, so there is no difference in inference efficiency. Importantly, our method enables fine-tuning of a publicly available 7B model within three hours. This demonstrates the practicality and accessibility of our approach.

\begin{table}[ht]
    \centering
    \caption{Training efficiency comparison of five methods evaluated on DeepSeek-Coder with our benchmark.}
    \label{tab:efficiency_comparison}
    \resizebox{\linewidth}{!}{%
        \begin{tabular}{l c c}
            \toprule
            \textbf{Method} & \textbf{Device} & 
            \textbf{Training Time}\\
            \midrule
            Instruct Tuning & A6000 (48GB) $\times1$ & 20\,min/epoch $\times3$ \\
            PromptMit & A6000 (48GB) $\times1$ & - \\
            SVEN & A6000 (48GB) $\times1$ & 60\,min/epoch $\times3$ \\
            DPO & A6000 (48GB) $\times1$ & 45\,min/epoch $\times3$ \\
            \textbf{Ours} & A6000 (48GB) $\times1$ &  49\,min/epoch $\times3$ \\
            \bottomrule
    \end{tabular}}
\end{table}


\RQSummary{RQ2}{Our proposed method consistently outperforms the evaluated baselines in compliance with the targeted non-functional quality criteria, achieving relative QRS improvements of up to 45.0\% on certain model–benchmark combinations while maintaining functional correctness.}

\subsection{RQ3: How do the individual components and design choices of our method affect performance?}
\label{Sec:RQ3}

Our method jointly optimizes three objectives, namely the language modeling loss $\mathcal{L}_{\mathrm{LM}}$, ranking loss $\mathcal{L}_{\mathrm{rank}}$, and KL-divergence loss $\mathcal{L}_{\mathrm{KL}}$, together with adaptive token-level weighting. These components aim to improve non-functional code quality while preserving functional correctness (Section~\ref{our_method}). To examine their individual contributions, we construct four controlled variants: \textbf{NoLM}, which removes $\mathcal{L}_{\mathrm{LM}}$; \textbf{NoRank}, which removes $\mathcal{L}_{\mathrm{rank}}$; \textbf{NoKL}, which removes $\mathcal{L}_{\mathrm{KL}}$; and \textbf{UniformWeight}, which assigns a constant weight of 1 to every token instead of using adaptive token weights. We compare these variants with the complete method using DeepSeek-Coder on \textit{MBPP-sanitized}, \textit{LiveCodeBench}, and \textit{CodeContests}.

\begin{figure*}[htbp]
    \centering
    \includegraphics[width=\linewidth]{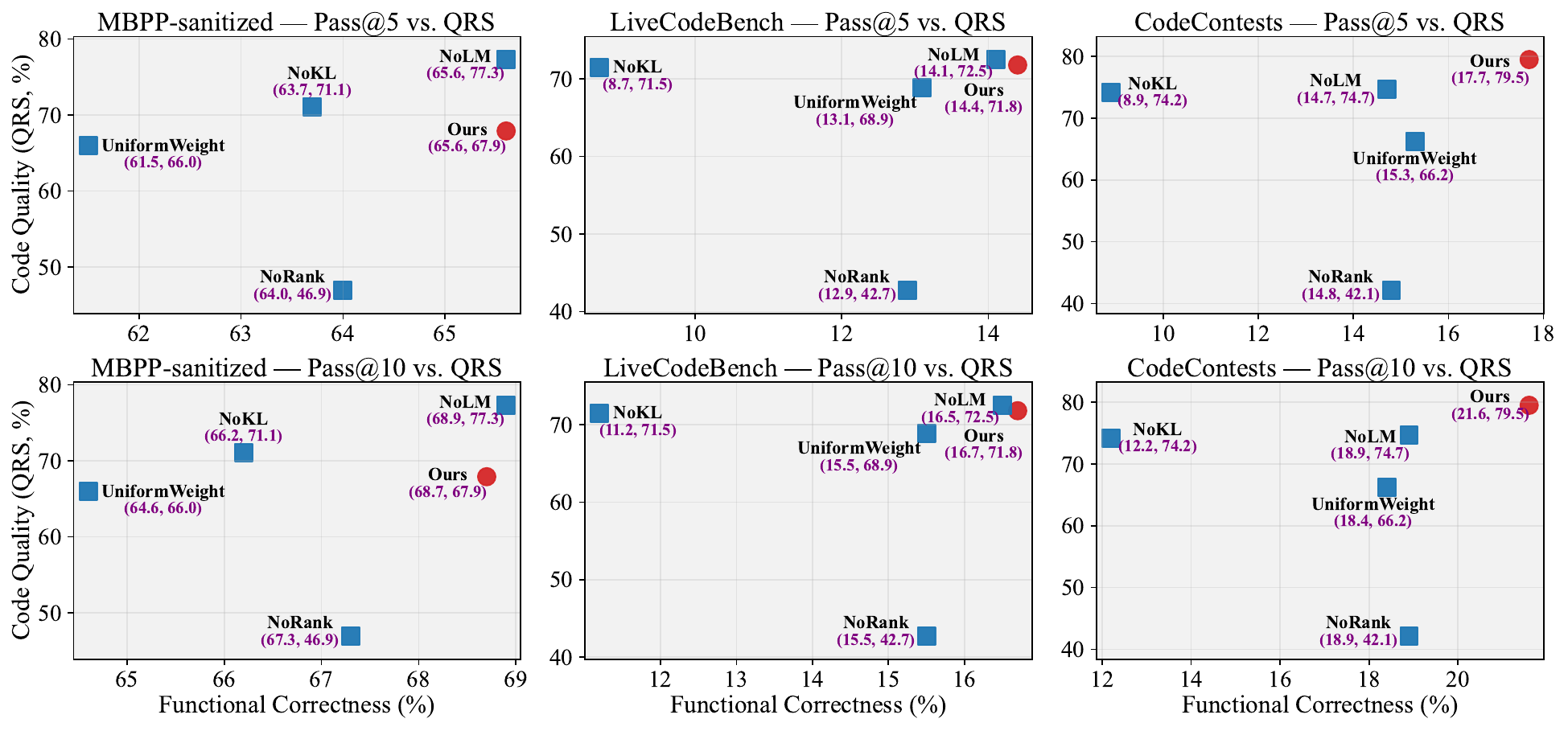}
    \caption{Trade-off between non-functional code quality (QRS) and functional correctness (Pass@k) for our complete method (red circles) and ablation variants (blue squares) across three benchmark datasets. Higher values on both axes indicate better performance.}
    \label{fig:abl_com}
\end{figure*}

Fig.~\ref{fig:abl_com} jointly presents non-functional code quality (QRS) and functional correctness (Pass@k) for the complete method and its ablation variants. Since improving non-functional quality at the cost of substantial functional degradation would limit practical utility, we evaluate each component by considering both dimensions rather than QRS alone.

\paragraph{Effect of $\mathcal{L}_{\mathrm{rank}}$.}
Removing $\mathcal{L}_{\mathrm{rank}}$ causes the most pronounced and consistent degradation in non-functional code quality. QRS decreases from 67.9\% to 46.9\% on \textit{MBPP-sanitized}, from 71.8\% to 42.7\% on \textit{LiveCodeBench}, and from 79.5\% to 42.1\% on \textit{CodeContests}, while the changes in Pass@k are comparatively smaller. These results demonstrate that $\mathcal{L}_{\mathrm{rank}}$ provides the key comparative supervision for distinguishing issue-compliant from issue-violating code.

\paragraph{Effect of $\mathcal{L}_{\mathrm{LM}}$.}
NoLM achieves a higher QRS on \textit{MBPP-sanitized} and a slightly higher QRS on \textit{LiveCodeBench}, but this advantage is not consistent across datasets or functional metrics. On \textit{CodeContests}, removing $\mathcal{L}_{\mathrm{LM}}$ decreases QRS from 79.5\% to 74.7\% and Pass@5 from 17.7\% to 14.7\%. This suggests that $\mathcal{L}_{\mathrm{LM}}$ complements quality-oriented optimization by helping retain the model's generation capability, particularly on more challenging tasks.

\paragraph{Effect of $\mathcal{L}_{\mathrm{KL}}$.}
Although NoKL maintains competitive QRS values, it exhibits noticeable degradation in functional correctness, particularly on \textit{LiveCodeBench} and \textit{CodeContests}. For example, on \textit{CodeContests}, Pass@5 decreases from 17.7\% to 8.9\%, and Pass@10 decreases from 21.6\% to 12.2\%. This supports the role of $\mathcal{L}_{\mathrm{KL}}$ in limiting excessive deviation from the original model and preserving functional behavior during quality optimization.

\paragraph{Effect of Adaptive Token Weighting.}
Compared with UniformWeight, adaptive token weighting consistently improves QRS across all three benchmarks, from 66.0\% to 67.9\% on \textit{MBPP-sanitized}, from 68.9\% to 71.8\% on \textit{LiveCodeBench}, and from 66.2\% to 79.5\% on \textit{CodeContests}. The improvement is particularly pronounced on \textit{CodeContests}, where QRS increases by approximately 20.1\%, from 66.2\% to 79.5\%. Meanwhile, Pass@5 increases from 15.5\% to 17.7\% and Pass@10 from 18.4\% to 21.6\%. These results show that emphasizing quality-relevant tokens provides more effective supervision than weighting all tokens equally, with particularly clear benefits on more challenging code-generation tasks.


Overall, the components play complementary roles. $\mathcal{L}_{\mathrm{rank}}$ provides the primary comparative signal for improving non-functional quality, while adaptive token weighting strengthens this signal by emphasizing quality-relevant tokens. $\mathcal{L}_{\mathrm{LM}}$ and $\mathcal{L}_{\mathrm{KL}}$ help retain the model's generation capability and functional correctness during optimization.

\RQSummary{RQ3}{The ablation results demonstrate that the proposed components complement each other in improving non-functional code quality while preserving functional correctness.}

\subsection{RQ4: To what extent do users prefer code from our optimized model over the baseline?}

In RQ1–RQ3, we demonstrated that our method effectively improves the non-functional quality of code generated by LLMs while maintaining functional correctness. In this RQ, we conduct a human evaluation to assess whether the optimized code is perceived as higher quality than baseline outputs, particularly regarding adherence to coding standards and best practices. Specifically, we use the original DeepSeek-Coder as the baseline and compare it with DeepSeek-Coder optimized using our method.

To this end, we randomly sample 100 tasks from three benchmarks, ensuring that each task produces valid (compilable and runnable) code from both the baseline and optimized models. For each task, we randomly sample one valid solution from each model, resulting in 100 code pairs (200 programs in total). To ensure unbiased evaluation, the positions of the two solutions within each pair are randomized. Eight reviewers, including four graduate students and four professional developers with over five years of Python experience, perform a blind review. Before evaluation, participants receive a brief refresher on coding standards, code smells, and relevant Python Enhancement Proposals (PEPs). Each pair is then presented anonymously, and reviewers label one of three outcomes: \emph{solution A is higher quality}, \emph{solution B is higher quality}, or \emph{no clear difference}.

\begin{figure}[htbp]
    \centering
    \includegraphics[width=\linewidth]{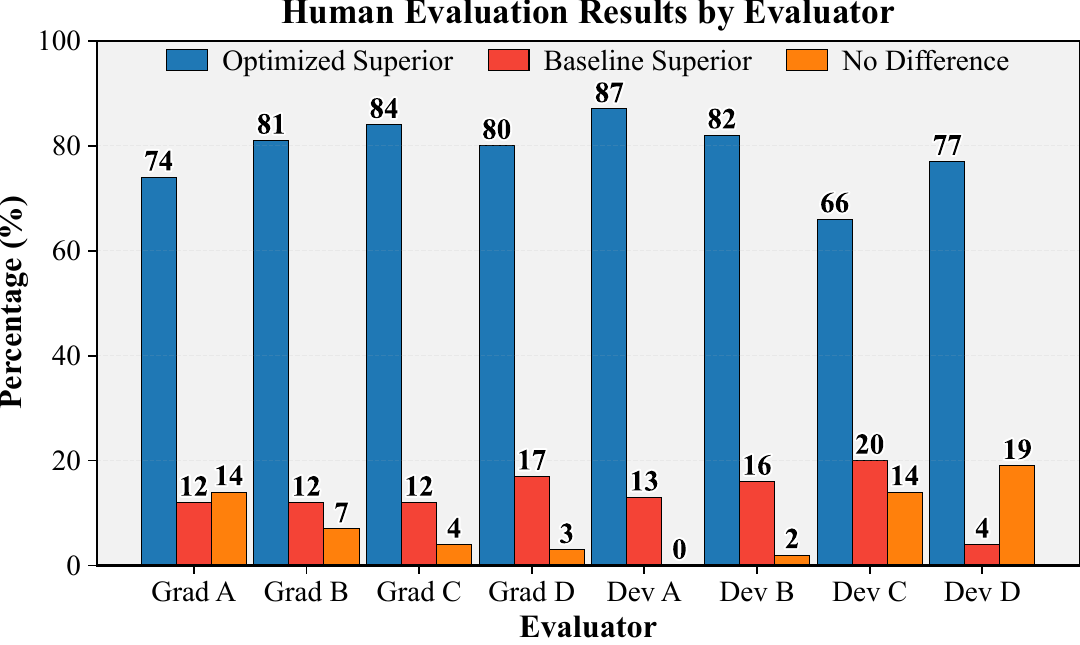}
    \caption{Human evaluation results on 100 code pairs. Each bar (blue/red/orange) shows the percentage of evaluators who favored the optimized model, the baseline model, or found no clear difference.}
    \label{fig:human_eval_comparison}
\end{figure}

Fig.~\ref{fig:human_eval_comparison} shows the distribution of reviewer decisions across the 100 code pairs. On average, 78.9\% of the pairs are judged to favor the optimized model, 13.3\% favor the baseline, and 7.9\% show no clear difference. We also report Agreement (\%), defined as the percentage of tasks where an evaluator’s judgment aligns with the majority vote~\cite{artstein2008inter}. An average Agreement of 82.9\% indicates high consistency among evaluators, supporting the reliability of our findings.
These results suggest that the optimized model tends to produce higher-quality code, as perceived by human evaluators.

\RQSummary{RQ4}{The user study demonstrates a clear human preference for the outputs of our optimized model, thereby reinforcing the practical effectiveness of our approach.}

\section{Discussion}
\subsection{Does Code Size Affect the Observed QRS Improvements?}
\label{sec:discussion_qrs_size}

QRS is computed from the number and severity of quality issues detected in a generated program. One potential concern is that longer programs may naturally provide more opportunities for static analysis tools to identify quality issues. Consequently, an improvement in QRS could potentially arise if the optimized model systematically generates shorter programs rather than code with better non-functional quality. We therefore examine whether the QRS improvements reported in RQ1 are accompanied by systematic reductions in generated code size.

We conduct this analysis on DeepSeek-Coder and restrict the comparison to programs that pass the complete functional test suite. This restriction allows us to compare the non-functional quality of solutions that already satisfy the required functionality. For these programs, we report QRS and Pass Rate together with two measures of code size: the average number of tokens and the average number of Python AST statement nodes. Token counts are computed using the DeepSeek-Coder tokenizer, while statement counts include Python AST statement nodes from successfully parsed programs.

Table~\ref{tab:rq1_correct_quality_deepseek} presents the results. The optimized model consistently achieves higher QRS and Pass Rate across all three benchmarks. On \textit{MBPP-sanitized}, for example, QRS increases from 46.2\% to 70.0\%, while Pass Rate increases from 3.3\% to 42.1\%. However, this improvement is not accompanied by a reduction in code size. The average token count changes only slightly from 85.8 to 86.0, while the average number of statements increases from 6.9 to 7.1. A similar pattern is observed on \textit{LiveCodeBench}, where the optimized solutions have both a higher average token count (88.7 versus 84.9) and a higher average statement count (8.1 versus 7.9), despite achieving a substantially higher QRS of 94.9\% compared with 73.9\%.

\begin{table}[ht]
    \centering
    \caption{Non-functional quality and code-size statistics of functionally correct programs generated by DeepSeek-Coder. QRS and Pass Rate are computed only over programs that pass the complete functional test suite.}
    \label{tab:rq1_correct_quality_deepseek}
    \resizebox{\linewidth}{!}{%
    \begin{tabular}{llrrrr}
        \toprule
        \textbf{Dataset} & \textbf{Approach} & \textbf{QRS} & \textbf{Pass Rate} & \textbf{Avg. Tokens} & \textbf{Avg. Statements} \\
        \midrule
        \multirow{2}{*}{MBPP-sanitized}
        & Original  & 46.2\% & 3.3\%  & 85.8 & 6.9 \\
        & Optimized & \textbf{70.0\%} & \textbf{42.1\%} & 86.0 & 7.1 \\
        \midrule
        \multirow{2}{*}{LiveCodeBench}
        & Original  & 73.9\% & 55.1\% & 84.9 & 7.9 \\
        & Optimized & \textbf{94.9\%} & \textbf{90.6\%} & 88.7 & 8.1 \\
        \midrule
        \multirow{2}{*}{CodeContests}
        & Original  & 50.2\% & 14.9\% & 154.4 & 14.4 \\
        & Optimized & \textbf{90.9\%} & \textbf{83.0\%} & 136.3 & 12.7 \\
        \bottomrule
    \end{tabular}%
    }
\end{table}

On \textit{CodeContests}, the optimized programs are shorter on average, with the token count decreasing from 154.4 to 136.3 and the statement count decreasing from 14.4 to 12.7. Nevertheless, this reduction is not observed on the other two benchmarks, where optimized programs are similar in size or slightly larger. Therefore, the optimized model does not exhibit a systematic tendency to generate shorter programs across datasets. More importantly, substantial QRS improvements remain evident on \textit{MBPP-sanitized} and \textit{LiveCodeBench} even when the optimized programs are not shorter.

These observations provide evidence that the QRS improvements cannot be explained solely by reductions in generated code size. Instead, the optimized model achieves higher non-functional quality even among functionally correct solutions of comparable or greater size. 

\subsection{Does Using the Same Quality Criteria for Training and Evaluation Introduce Circularity?}
\label{sec:discussion_evaluation_criteria}

Our dataset construction and evaluation use the same unified quality criteria derived from \textsc{Pylint}, \textsc{Flake8}, and \textsc{SonarQube}. This design may raise a concern that evaluating QRS and Pass Rate using these criteria favors the optimized models because they are also involved in constructing the training data. However, the primary objective of our evaluation is to measure \emph{targeted effectiveness}: whether the optimized model learns to avoid the predefined non-functional quality issues explicitly targeted during training. Static-analysis rules provide explicit and reproducible definitions of these issues and have been widely used to evaluate the non-functional quality of LLM-generated code. Prior studies employ \textsc{Pylint}~\cite{liu2024refining,siddiq2022empirical,siddiq2024quality,rabbi2024ai}, \textsc{Flake8}~\cite{feng2023investigating}, and \textsc{SonarQube}~\cite{yeticstiren2023evaluating} for this purpose. Our study integrates the rules from all three tools to provide broader coverage. Under these predefined criteria, a reduction in detected violations directly reflects improved compliance with the quality properties targeted by our method.

Importantly, the use of the same criteria for optimization and evaluation should be distinguished from evaluating on the same code samples. Although the quality criteria are fixed, our evaluation is conducted on independent benchmark tasks from \textit{MBPP-sanitized}, \textit{LiveCodeBench}, and \textit{CodeContests}, which are separate from the APPS tasks used to construct the training pairs. The consistent improvements across these benchmarks provide evidence that the model learns transferable preferences for the targeted quality issues rather than merely reproducing repairs for the training programs.

Holding out quality criteria would instead evaluate \emph{criteria-level generalization}, which is fundamentally different from the task-level generalization considered above. Different quality issues can involve substantially different code patterns and repair transformations. If a particular issue and its corresponding repair pattern are absent from the quality training data, our training procedure provides no explicit comparative supervision from which to learn that specific quality preference. For example, training pairs illustrating R0912 (\textit{too-many-branches}) teach the model preferences related to reducing excessive control-flow branching, but they do not directly provide the repair signal required for R1705 (\textit{no-else-return}), which concerns a different structural pattern. This is analogous to supervised learning more generally, where generalization to new instances of learned concepts is distinct from recognizing entirely unseen target concepts.

Our evaluation protocol therefore focuses on whether the model generalizes the learned quality preferences to unseen tasks and programs under the targeted criteria, which directly corresponds to the objective of our method. QRS and Pass Rate quantify this targeted improvement using the same explicit criteria that define the quality properties being optimized. 
Evaluating transfer to held-out rules or additional analyzers would instead assess \emph{criteria-level generalization}, which constitutes a distinct research question beyond the scope of the present study.

\section{Related work}
\label{sec:qua_issues}
Most studies on improving code quality focus on functional correctness, typically measured by the percentage of generated programs that pass unit tests~\cite{liu2024refining}. To this end, RL–based methods such as CodeRL~\cite{le2022coderl}, PPOCoder~\cite{shojaee2023execution}, and CompCoder~\cite{wang2022compilable} incorporate feedback from tests or compilers to optimize models. Other approaches improve correctness by tuning generation parameters such as temperature~\cite{blyth2024creative,arora2024optimizing}. However, passing unit tests does not fully ensure high-quality code~\cite{liu2024refining, velasco2025causal}. Evidence shows that code style, maintainability, and other non-functional quality issues~\cite{siddiq2022empirical,siddiq2024quality} remain significant concerns in LLM-generated programs~\cite{liu2024refining}.

Numerous studies~\cite{siddiq2022empirical,siddiq2024quality,liu2024refining} have investigated non-functional quality issues in LLM-generated code. For instance, Liu et al.~\cite{liu2024refining} report that even when ChatGPT-generated programs pass all test cases, up to 53\% of Java and 37\% of Python outputs suffer from poor stylistic or limited maintainability.
Similar findings are reported in~\cite{moratis2024write,siddiq2024quality}, and Rabbi et al.~\cite{rabbi2024ai} note that modifications to ChatGPT-generated code may introduce new quality issues. Feng et al.~\cite{feng2023investigating} identify frequent pycodestyle violations in ChatGPT-generated Python code using crowdsourced annotations.
Yetiştiren et al.~\cite{yeticstiren2023evaluating} report maintainability concerns across multiple LLMs (e.g., Copilot~\cite{copilot2021github}, CodeWhisperer~\cite{amazon2023codewhisperer}, and ChatGPT~\cite{achiam2023gpt}), with resolving code smells taking 5.6–9.1 minutes.
Clark et al.~\cite{clark2024quantitative} assess LLM code complexity via Halstead metrics,  while Siddiq et al.~\cite{siddiq2022empirical} further investigate that quality flaws in training data may propagate into model outputs.

Few studies have aimed to improve the non-functional quality of LLM-generated code. Velasco et al.~\cite{velasco2025causal} encourage criteria-compliant generation by prompting models to avoid code smells. Yao et al.~\cite{yao2025training} propose REAL, which uses MyPy-derived program-analysis feedback to optimize maintainability-oriented properties through reinforcement learning. REAL therefore studies a different optimization setting centered on MyPy-specific feedback, whereas our work investigates preference learning over paired violating–compliant programs defined by a unified set of quality criteria from multiple static analyzers. Relatedly, Naik et al.~\cite{naik2025metalint} propose MetaLint, an instruction-following framework that uses rejection-sampling DPO with linter-derived rewards to detect and localize best-practice violations, rather than improving the quality of generated code.


Unlike previous methods, we propose a unified framework combining dataset construction, adaptive token weighting, and ranking loss for quality-aware preference learning. Our method improves non-functional quality more effectively than baselines while preserving functional correctness.

\section{Threats to validity}
\label{limitation}

\paragraph{Threats to Internal Validity}
Internal validity concerns whether the results are unbiased. 
To mitigate randomness in individual outputs, we first compute the QRS and Pass Rate across all generated solutions for each task, and then average these scores across all tasks, which reduces noise from any single sample. To minimize potential bias introduced by tool-specific behavior, we use a unified set of 527 rules aggregated from three widely used static analyzers (\textsc{Pylint}~\cite{pylint}, \textsc{Flake8}~\cite{ziade27flake8}, and \textsc{SonarQube}~\cite{sonarqube2024}), which offer diverse yet complementary coverage of non-functional quality issues. 
Nevertheless, these analyzers have inherent limitations and may not fully capture all aspects of non-functional code quality, potentially introducing residual bias.
Exploring additional analyzers and more advanced quality assessment techniques remains an important direction. 
In addition, we evaluate on two code LLMs (DeepSeek-Coder and Qwen2.5-Coder) to reduce model-specific bias. While generation may still be affected by configurations and runtime environments, we use consistent parameters and environments for both the original and optimized models. Future work will explore robustness under broader deployment settings.


\paragraph{Threats to External Validity}
External validity concerns the generalizability of our findings. Most prior studies rely on a single static analysis tool (e.g., \textsc{Pylint}, \textsc{Flake8}, or \textsc{SonarQube}) to assess non-functional code quality, which may provide incomplete coverage of quality issues. Although our study integrates three complementary tools and consolidates 527 detection rules covering a broad range of detectable non-functional quality issues, some quality aspects may remain beyond the scope of these static analyzers.
Our study focuses on whether the learned quality preferences generalize to unseen programming tasks under the predefined quality criteria. We therefore evaluate our method on three public benchmarks with different task distributions. The consistent improvements across these benchmarks provide evidence of task-level generalizability under the considered criteria. Generalization to quality criteria or static analyzers beyond those considered in this study remains a direction for future research.
In addition, our evaluation is limited to two open-source LLMs with up to 7B parameters due to computational constraints. Although the three benchmarks cover diverse programming tasks, they may not fully represent real-world development scenarios. Future work could further evaluate our framework on larger models, more diverse development tasks, and broader quality criteria.

\section{Conclusion}
\label{conclusion}
In this paper, we propose a quality-aware comparative optimization framework to improve the non-functional quality of code generated by LLMs. The core of our approach is the construction of paired criteria-violating and -compliant samples, the introduction of adaptive token weighting to emphasize quality-sensitive regions, and a comparative ranking objective for effective preference learning. Experiments on three benchmarks demonstrate that our method substantially improves compliance with the targeted non-functional quality criteria, while preserving functional correctness. Future work will extend this framework to additional programming languages, more complex real-world projects, and a broader set of quality rules and static analysis standards.

\ifCLASSOPTIONcompsoc
  \section*{Acknowledgments}
\else
  \section*{Acknowledgment}
\fi

This work was supported by the National Natural Science Foundation of China (Grant Nos. 62302125, 62272132, U25B2052).

\ifCLASSOPTIONcaptionsoff
   \newpage
\fi



\bibliographystyle{IEEEtran}
\bibliography{reference}

\end{document}